\documentclass[sigconf, nonacm]{acmart}

\AtBeginDocument{%
  \providecommand\BibTeX{{%
    \normalfont B\kern-0.5em{\scshape i\kern-0.25em b}\kern-0.8em\TeX}}}

\newcommand\vldbdoi{XX.XX/XXX.XX}
\newcommand\vldbpages{XXX-XXX}
\newcommand\vldbvolume{14}
\newcommand\vldbissue{1}
\newcommand\vldbyear{2020}
\newcommand\vldbauthors{\authors}
\newcommand\vldbtitle{\shorttitle} 
\newcommand\vldbavailabilityurl{URL_TO_YOUR_ARTIFACTS}
\newcommand\vldbpagestyle{plain} 

\usepackage{graphicx}
\usepackage{balance}
\usepackage{xcolor}
\usepackage[most]{tcolorbox}
\usepackage{amsthm}
\usepackage{bbm}
\usepackage{subfigure}
\usepackage{url}
\usepackage{booktabs}
\usepackage{enumitem}
\usepackage{booktabs}
\usepackage{caption}
\usepackage{makecell}
\usepackage{multirow}
\usepackage{tcolorbox}
\usepackage{pifont}
\usepackage{xspace}
\newcommand{\sys}{\textsc{EMIT}\xspace}

\newfloat{figtab}{htb}{fgtb}
\makeatletter
  \newcommand\figcaption{\def\@captype{figure}\caption}
  \newcommand\tabcaption{\def\@captype{table}\caption}
\makeatother

\usepackage{enumitem}
\usepackage{pifont}
\usepackage[ruled, vlined, linesnumbered]{algorithm2e}
\usepackage{amsmath}
\usepackage{booktabs}
\usepackage{multirow}

\usepackage{adjustbox}
\usepackage{tabularx}
\usepackage{graphicx}
\usepackage{float}
\usepackage{xcolor}
\SetKwInOut{Parameter}{Parameters}
\usepackage{hyperref}
\def\Snospace~{Section {}}

\usepackage{xspace}

\begin{document}

\title{EMIT: Micro-Invasive Database Configuration Tuning}

\author{Jian Geng}
\affiliation{%
  \institution{Harbin Institute of Technology}
}
\email{gengj@stu.hit.edu.cn}

\author{Hongzhi Wang}
\affiliation{%
  \institution{Harbin Institute of Technology}
}
\email{wangzh@hit.edu.cn}

\author{Yu Yan}
\affiliation{%
  \institution{Harbin Institute of Technology}
}
\email{yuyan@hit.edu.cn}


\sloppy
\fancyhead{} 

\begin{abstract} 
The process of database knob tuning has always been a challenging task. Database Administrators (DBAs) typically rely on their extensive experience and domain knowledge to tune the systems, which is a laborious and time-consuming endeavor. Recently, database knob tuning methods has emerged as a promising solution to mitigate these issues. However, these methods still face certain limitations.On one hand, when applying knob tuning algorithms to optimize databases in practice, it either requires frequent updates to the database or necessitates acquiring database workload and optimizing through workload replay. The former approach involves constant exploration and updating of database configurations, inevitably leading to a decline in database performance during optimization. The latter, on the other hand, requires the acquisition of workload data, which could lead to data leakage issues. Moreover, the hyperparameter configuration space for database knobs is vast, making it challenging for optimizers to converge. These factors significantly hinder the practical implementation of database tuning.

To address these concerns, we proposes an efficient and micro-invasive knob tuning method. This method relies on workload synthesis on cloned databases to simulate the workload that needs tuning, thus minimizing the intrusion on the database. And we utilizing a configuration replacement strategy to filter configuration candidates that perform well under the synthesized workload to find best configuration. And during the tuning process, we employ a knowledge transfer method to extract a common high-performance space, to boost the convergence of the optimizer. Experiment shows that  our method can achieve the same level of $0.9\times$ best performance with $1.8\times$ to $12.5 \times$ fewer iterations. End to end performance test further confirm that our system is effective for configuration tuning while doing minimal instruction to the database.

\end{abstract}
\maketitle

\pagestyle{\vldbpagestyle}
\begingroup\small\noindent\raggedright\textbf{PVLDB Reference Format:}\\
\vldbauthors. \vldbtitle. PVLDB, \vldbvolume(\vldbissue): \vldbpages, \vldbyear.\\
\href{https://doi.org/\vldbdoi}{doi:\vldbdoi}
\endgroup
\begingroup
\renewcommand\thefootnote{}\footnote{\noindent
This work is licensed under the Creative Commons BY-NC-ND 4.0 International License. Visit \url{https://creativecommons.org/licenses/by-nc-nd/4.0/} to view a copy of this license. For any use beyond those covered by this license, obtain permission by emailing \href{mailto:info@vldb.org}{info@vldb.org}. Copyright is held by the owner/author(s). Publication rights licensed to the VLDB Endowment. \\
\raggedright Proceedings of the VLDB Endowment, Vol. \vldbvolume, No. \vldbissue\ %
ISSN 2150-8097. \\
\href{https://doi.org/\vldbdoi}{doi:\vldbdoi} \\
}\addtocounter{footnote}{-1}\endgroup

\ifdefempty{\vldbavailabilityurl}{}{
\vspace{.3cm}
\begingroup\small\noindent\raggedright\textbf{PVLDB Artifact Availability:}\\
The source code, data, and/or other artifacts have been made available at \url{\vldbavailabilityurl}.
\endgroup
}

\section{Introduction}

With the development of big data and cloud databases, it has become increasingly essential to provide better services by adjusting the database knob to improve its performance.
However, modern databases usually have hundreds of configurable knobs ~\cite{dias_automatic_nodate,best_postgres} and complex influence relationships among the knobs, making database knob tuning an NP-hard problem ~\cite{sullivan_using_2004}. In practice, database administrators need to spend a considerable amount of time finding a configuration that is suitable to a specific workload\cite{debnath2008sard}. This requires database administrators to make use of their historical experiences, coupled with a tedious and mechanized tuning process to achieve this goal~\cite{van_aken_automatic_2017}.
In order to solve the above problems, many machine learning based methods, such as the Bayesian Optimization based~\cite{shen_rover_2023,towards_safe,llama_tune} and the Reinforcement Learning based~\cite{cdb_tune, hunter} methods have been proposed for knob tuning, and have achieved remarkable performance\cite{zhao2023automatic, zhang_facilitating_2022}. Additionally leveraging historical experience for knob tuning is critical, which enables the optimizer to optimize knobs with the history knowledge instead of exploring from zero. 

To facilitate knowledge transfer, OtterTune~\cite{van_aken_automatic_2017} adopts \textsc{Workload Mapping} strategy to identify the most similar historical experience and transfer its knowledge. And recent research  ~\cite{shen_rover_2023, hunter} proposes the use of meta-learning method~\cite{NEURIPS2018_14c879f3}, which are based on the Kendall distance metric to assign appropriate weights to historical experiences for knowledge transfer, and have achieved impressive results.


\noindent\textbf{\underline{Motivations.}} Despite some recent studies demonstrating promising experimental outcomes, when applying these database knob tuning methods to actual database systems, they usually have a \textbf{large intrusion} into the database: \textbf{i).} frequent changes to the database knobs to continually explore more optimal database configurations, or \textbf{ii).} obtaining users' database workloads and tuning the knobs through workload replay. The former can lead to unstable database performance during the tuning period, causing significant performance fluctuations for client applications. The latter, however, requires access to sensitive workload information, which could potentially lead to data leakage issues. Current tuning methods can't take into account both aspects well, making it challenging for knob tuning algorithms to be successfully implemented. Therefore, it is very necessary to develop a knob tuning system that is minimally invasive to database.


In order to avoid database performance fluctuations or sensitive workload information collection as much as possible while tuning the database efficiently, we have summarized the challenges that need to be solved as follows:

i). In order to avoid large performance fluctuations on the database and avoid obtaining workload to prevent data leakage. It will be no longer feasible to directly use optimization algorithms to tune.  \textbf{(C1)}

ii). If C1 can be solved, and the optimizer recommends some configuration candidates. However, since we cannot obtain a relationship between database workload performance and the configuration, how do we efficiently locate the suitable candidate configuration while having as little impact on the database as possible?  \textbf{(C2)}

iii). The knowledge of history tasks contains sufficient information helpful in tuning for new scenarios. However, current studies~\cite{zhang_facilitating_2022, ResTune, QTune, watuning, shen_rover_2023, van_aken_automatic_2017}  all usually need to deploy a number of configurations to obtain their corresponding performances, in order to 'correct' the learned knowledge for new workload, which may also cause database performance fluctuations.\textbf{(C3)}

The challenges rises for the need that, the optimizer has to frequently explore both good and bad (configurations that improve database performance and otherwise) configurations to construct the correlation distribution between database knobs and performance, which inevitably leads to significant intrusion on databases. Motivated by this, we propose \sys, an  \underline{E}fficient and \underline{M}icro-\underline{I}nvasive knob \underline{T}uning approach. To achieve micro-invasiveness, \sys employs a cloned database to simulate workload execution, making the cloned database behave similarly to the target database requiring tuning. This ensures that the optimization outcomes are applicable to the database requiring tuning without frequently updating database knobs. In the tuning process, we propose a common knowledge transfer approach, to fully exploit the knowledge contained in historical experience and accelerate the tuning process. Next we introduce how our method addresses the challenges.

\noindent\textbf{\underline{Our Methods.}} i) For C1, we have designed a workload synthesis method to avoid collecting sensitive information. Motivated by ~\cite{van_aken_automatic_2017}, we utilize the DBMS's runtime metrics to characterize the behavior of the target workload. We then synthesize a workload using existing basic workloads to make the synthesized workload as similar as possible to the database-executed workload at the runtime metric level. To address C2, we have designed a Recursive Configuration Selection algorithm to select and deploy promising configurations to the production database based on the synthesized workload, thus achieving performance improvement with relatively few iterations. During the optimization period, to prune the search space of the optimizer (C3) and boost the tuning convergence of database knobs, we propose a Common Knowledge Extract algorithm. This algorithm identifies those promising knob configuration spaces, and combines information obtained in the algorithm with the modified version of SMAC Bayesian optimization algorithm to accelerate the tuning process.

\noindent\textbf{\underline{Contributions.}}Based on the method we propose, our contributions can be summarized as follows:

i). We introduce a workload synthesis method that can simulate the behavior of workloads to be optimized at the runtime metric level.

ii). We propose a recursive configuration selection algorithm, which transfers optimal configurations based on synthetic workloads to the database to be tuned, achieving performance improvements with fewer steps.

iii). We present a common knowledge transfer method for discovering promising parameter spaces, thereby pruning unnecessary space exploration.

iv). We introduce a modified version of the SMAC algorithm that utilizes common knowledge to guide its exploration process in the early stages, enabling the optimizer to converge more effectively.

The remaining sections of the paper are organized as follows: We first introduce some preliminaries in ~\autoref{sec:preliminaries}. An overview of \sys, its overall structure and components are presented in ~\autoref{sec:overview}. The synthesis of workloads is discussed in ~\autoref{sec:workload}. Then we will describe how to accelerate the tuning process using historical experience in ~\autoref{sec:tuning}. And a configuration selection algorithm is introduced in ~\autoref{sec:strategy}. Finally, the tuning experimental results of our system are presented in ~\autoref{sec:experiment}, and the paper concludes with ~\autoref{sec:conclusion}. Note that in the following sections, knob tuning and configuration tuning both refer to the optimization of tunable hyperparameters in a database. Therefore, we will use these two terms interchangeably.

\section{RELATED WORK}\label{sec:related}

\noindent\textbf{Configuration Tuning:} In our review of database configuration tuning methods, we categorize them into three types based on how they adjust the tuning knobs: Heuristic method, Bayesian Optimization (BO) Based method, and Reinforcement Learning (RL) based method: 

\textbf{i) Heuristic:} 
Sullivan et al. \cite{sullivan_using_2004} employ Probability Theory to determine whether a given set of tuning rules is likely to improve the target workload, making decisions on the best tuning actions based on the probability. PGTune\cite{PGTune} boosts the database configuration tuning process by leveraging rules summarized by experts. OpenTuner\cite{ansel2014opentuner} first sample initial points and then explore new points based on the observed performance, allocating more chances for exploring when good performance is observed and less chances otherwise. BestConfig\cite{bestconfig} optimizes the sampling stage with their proposed Divide and Diverge Sampling method, enabling efficient sampling and exploration within a given budget. 
\textbf{ii) BO-based:} 
Bayesian optimization can find good configurations within a relatively small number of iterations in high-dimensional spaces \cite{inquiry, zhang_facilitating_2022}. 
Consequently, some studies for database knob tuning based on bayesian optimization have emerged: 
iTuned \cite{ibtune} has been proposed to adjust the buffer pool size of database instances in order to reduce memory utilization while ensuring performance. ResTune \cite{ResTune} goes beyond just tuning memory allocation, allowing for the tuning of any resource consumption by the database instance to decrease the hardware resource comsumption. To reduce the dimension of exploration space, LlamaTune \cite{llama_tune} employs random low-dimensional projections to narrow down the search space for tuning. Meanwhile, Rover \cite{shen_rover_2023} adopts expert rules to guide the exploration process of Bayesian optimization, enhancing the safety of the exploration. 
\textbf{iii) RL-based:}
The agent of RL explores promising configuration space with trial-and-error strategies, and finally convergent to a near optimal configuration.
CDBTune\cite{cdb_tune} employs the deep deterministic policy gradient method (DDPG) to optimize configurations, but this approach does not model the characteristics of workloads, making it less effective under dynamic workload. In contrast, QTune\cite{UDO} encodes workload queries, taking workload features into account during the optimization process. To accelerate RL, UDO\cite{UDO} categorizes knobs into 'light' and 'heavy' (those requiring a restart to take effect), thereby reducing the cost of updating 'heavy' knobs. Hunter\cite{hunter} utilizes a genetic algorithm explore efficiently in the early stage.

\noindent\textbf{Knowledge Transfer:} 
Database configuration tuning processes can be time-consuming\cite{debnath2008sard}. However, the experience gained from previous tuning tasks for similar workloads can be leveraged\cite{van_aken_automatic_2017}. Several studies have proposed methods for knowledge transfer in database knob tuning to fully utilize historical tuning experiences instead of starting from scratch each time. OtterTune\cite{van_aken_automatic_2017} uses a workload mapping approach, migrating the most similar historical experiences to the current workload based on observed workload characteristics. ResTune\cite{ResTune} and Rover\cite{shen_rover_2023} utilize the ratio of concordant pairs to weight historical experiences, producing an ensembled model that guides the tuning in new scenarios. QTune\cite{QTune} and WATuning\cite{watuning} can adapt to new tuning scenarios through fine-tuning of the model.

\noindent\textbf{Workload Synthesis:}
The work closest to ours is Stitcher\cite{stitcher}, which divides open-source workloads into different workload pieces and then uses machine learning models to combine different workload pieces to synthesize a workload behave similarly to the target workload. However, this method requires a significant amount of time to retrain the model when facing new hardware scenarios. Morfonios et al.\cite{morfonios2011consistent} and Galanis et al.\cite{galanis2008oracle} obtain the workloads by tracking user workloads and replay them, causing significant costs and need to access sensitive data. Kim et al.\cite{kim2014workload} directly capture hardware performance and statistical information from the OS-kernel level for workload synthesis, but its application is not in the field of database tuning. To the best of our knowledge, it lacks research on synthesizing database workloads, and there is currently no well solution.
\section{PRELIMINARIES}\label{sec:preliminaries}

\subsection{Problem Definition}

\noindent\textbf{Simple Configuration Tuning (SCT).} Consider a DBMS with $n$ knobs: 
$\{\theta_1, \theta_2, ..., \theta_n \}$. $\theta_i$ represents the $i$ th knob's configurable space, which could be continuous or discrete. Thus, the whole configuration space can be written as 
$\Theta = \theta_1 \times \theta_2 \times ... \times \theta_n$. The environment space $\Omega$ denotes the non-sensitive environmental information collected during database's workload execution. The database's performance metric to be optimized is denoted as objective function $f$, e.g. throughput, 95\%percentile latency and etc.
Given a workload $\omega$, simple configuration tuning aims to find a configuration $\theta^* \in \Theta$, where:
\begin{equation}
\begin{aligned} \label{eq:simpleTune}
& \theta^* = \mathop{\arg\max}_{\theta \in \Theta} f(\theta | \omega), 
\omega \in \Omega \\
\end{aligned}
\end{equation}

\noindent\textbf{Experience Enhanced Tuning (EET).} It is assumed that some historical tuning experiences have been collected. We note a historical trace as $e$, composed of historical configurations and their corresponding performance under a specific workload: $e=\{(config_1, perf_1), (config_2, perf_2), ...\}$. So the Experience Repository can be presented as $\mathcal{E} = \{ (e_1, \omega_1), (e_2, \omega_2), ..., (e_k, \omega_k),\}$. 

Given a target workload $\omega_t$, the objective of EET is to find a near optimal configuration $\theta^* \in \Theta$, with the information gained from $\mathcal{E}$. And in most cases, to achieve the same performance increase, the number of samples of EET should be much smaller than the number of samples of SCT.

\noindent\textbf{Workload Synthesis (WS)}. WS aims to synthesize a workload, so that it has characteristics similar to the original workload as much as possible. In this paper, we use different open source workload as basic workloads, e.g. YCSB~\cite{ycsb}, TPCC~\cite{tpcc} and etc. Thus given $k$ basic workloads $\omega_i$, and target workload $\omega_t$ to be synthesized, so the objective of WS is formulated as follows:

\begin{equation}
\begin{aligned}\label{alg:workloadSyn}
& \lambda^* = \mathop{\arg\max}_{\lambda \in \Lambda} \text{Similarity} \left( \sum_{i=1}^{k} \lambda_i \cdot \omega_i, \omega_t \right)  \\
& \text{s.t. } 0 \leq \lambda_i \leq 1 , \sum_{i=1}^{n} \lambda_i = 1, \omega \in \Omega
\end{aligned}
\end{equation}

\subsection{SMAC Optimization Algorithm}

Bayesian optimization has been proven to be an effective optimization method in hyperparameter tuning problem, as it can build an empirical Gaussian distribution with fewer sampling points to guide further exploration by the optimizer. SMAC~\cite{smac} is a variant of Bayesian optimization and has shown outstanding performance in database optimization problem~\cite{zhang_facilitating_2022}. Therefore, this paper adopts SMAC as the optimizer, and we will introduce Bayesian Optimization and the SMAC algorithm below.

\subsubsection{Bayesian Optimization}

Bayesian optimization is a powerful strategy for global optimization of black-box functions that are expensive to evaluate. It is particularly useful for optimizing hyperparameters of machine learning models where direct evaluation of the performance metric (e.g., validation accuracy) can be computationally expensive and time-consuming. Bayesian optimization iterates between two main steps: constructing a surrogate model to approximate the objective function, and using this model to determine the most promising points to evaluate next based on an acquisition function.

\begin{algorithm}[t]
\DontPrintSemicolon
\KwIn{Search Space $\mathcal{S}$, Max Iteration $\mathcal{I}$}
\KwOut{Best Point Explored $\mathcal{X}$}
\caption{Bayesian Optimization}\label{alg:bayesianOpt}

$initial\_points = \text{sample n points from } \mathcal{S}$

$\mathcal{P} = \{ (x_i, Observation(x_i)) | x_i \in initial\_points\}$

\While{$|\mathcal{P}| < \mathcal{I}$}{
    \text{// build surrogate model}\\
    $m = fit(\mathcal{P})$\\
    $p = p_i \text{ that maximize acquisition function } f(p_i, m)$\\
    \text{// get observation is time-consuming}\\
    $\mathcal{P} = \mathcal{P} \cup \{(p, Observation(p))\}$\\
    
}

\Return{\text{x with max observation }}.\;
\end{algorithm}

\subsubsection{\textit{SMAC Algorithm}}
SMAC is a variant of Bayesian optimization specifically designed to address the challenges posed by discrete parameter types in Gaussian Process regression. In the database configurations, some knobs can be of discrete or enumerated types, making SMAC an effective solution for such scenarios.

\textbf{Surrogate Model.} 
Unlike Vanilla Bayesian Optimization, which uses Gaussian Processes as surrogate models, Sequential Model-based Algorithm Configuration (SMAC) employs random forest as the surrogate model. Suppose we have a set of observed points $\mathcal{P} = \{(x_1, y_1), (x_2, y_2), \cdots, \{x_n, y_n)\}$. We initialize a random forest model to fit the data in $\mathcal{P}$. Given an unknown configuration $x$, we use each tree in the random forest to predict $x$, obtaining the mean and variance of the prediction.
Then, the mean and variance are used as inputs to the acquisition function, from which a score is computed. Subsequently, the unknown configuration with the highest score is selected as the point for the next observation.

\textbf{Acquisition Function.}
The acquisition function is used for balancing exploitation with exploration, where the optimizer uses it to score the next sampling point, thereby guiding the optimizer's subsequent explorations. The Expected Improvement (EI) function shown in ~\autoref{alg:EI} is commonly utilized in database tuning. It calculates the expected improvement of an unseen point over the current optimal point, and $p(y|x)$ denotes the probability given by surrogate model. In this paper, we will make modifications to the acquisition function in ~\autoref{sec:EESMAC}, aiming to facilitate knowledge transfer without affecting its convergency.

\begin{equation}
\begin{aligned}\label{alg:EI}
& EI(x) = \int_{-\infty}^{+\infty} max(y^* - y, 0)p(y|x)dy
\end{aligned}
\end{equation}

\begin{figure*}
\centering
\includegraphics[width=6in]{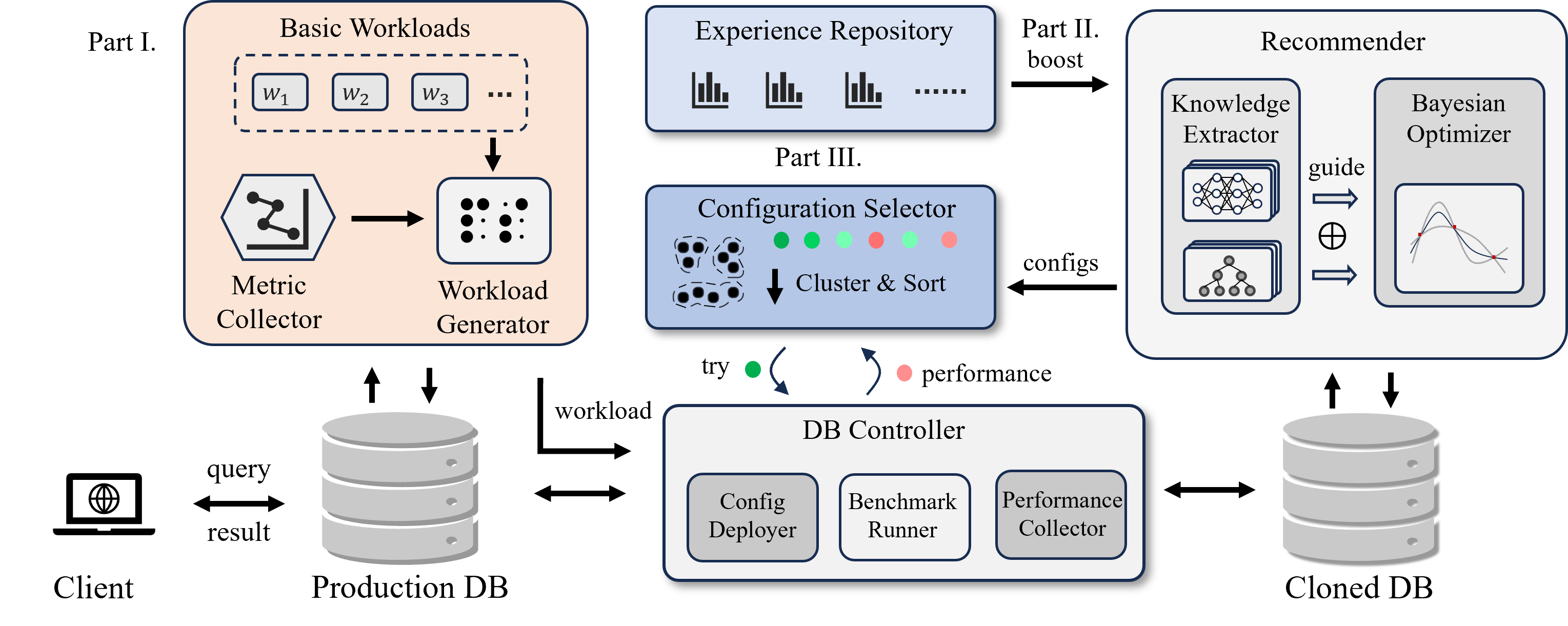}
\caption{Overview of EMIT}
\label{fig:overview}
\end{figure*}

\subsection{Key Term Definition}

In order to minimize redundancy in our writing and to make the paper easier to understand for readers, we have collect some key terms found within this whole paper that may be confusing, and we provide explanations for them below.
\begin{itemize}[leftmargin=*]
    \item \textbf{Production DB}: The database that is executing the workload. This database is the final target of the tuning algorithm, that is, the recommended optimal configuration will eventually be deployed on Production DB.
    \item \textbf{Cloned DB}: In order to simulate the behavior of runtime metric of Production DB, the Cloned DB is created that is exactly the same as Production DB at the software and hardware levels. This database is used to execute synthetic workloads, and the optimizer will recommend configurations based on the Cloned DB to avoid intrusion into the Production DB.
    \item \textbf{Experience Enhanced}: It refers to using the knowledge contained in historical experience to enhance the adaptability of algorithms to new scenarios.
    \item \textbf{Performance Regression}: A situation where the database still provides its service, but performs more slowly than before.
\end{itemize}

\section{OVERVIWE OF \sys}\label{sec:overview}

In this section, we will introduce the architectural composition of the entire system and its workflow in ~\autoref{fig:overview}. Since the tuning of database systems is a highly engineering-intensive task, therefore, this paper simplifies the problem in some aspects, so the assumptions and limitations involved will be detailed in ~\autoref{sec:assumption}.

\subsection{System Overview} \label{sec:system_overview}

The overview of the entire \sys system, as depicted in ~\autoref{fig:overview}, can be segmented into three main parts: i). Workload Synthesis, ii). Experience Enhanced Tuning, and iii). Configuration Replacement Strategy. 

For \textbf{part i}, we utilize a metric collector to gather runtime metrics from the Production DB. A workload generator then employs these metrics to compose basic workloads, yielding synthetic workloads. Subsequently, the DB Controller create a \textsc{Cloned DB} mirroring the hardware environment of the \textsc{Production DB}. The DB Controller executes the synthesized workload on the \textsc{Cloned DB} using a benchmark runner and gathers performance metrics using a Performance Collector. In \textbf{part ii}, we employ the recommender to optimize the performance of the \textsc{Cloned DB}. To expedite the efficiency of the optimizer exploration, the Recommender leverages historical experiences from the Experience Repository to extract historical knowledge to boost the convergence. Finally in \textbf{part iii}, once the Recommender suggests promising configuration settings, a Configuration Selector filters these candidate configurations, and the Config Deployer is then utilized to deploy configurations onto the \textsc{Production DB} until an appropriate configuration is identified.

\subsection{Assumptions}\label{sec:assumption}

Since database tuning is a complex engineering problem, this paper makes reasonable assumptions about the tuning tasks as follows instead of covering all aspects.

\textbf{i).} In the process of database knob tuning, some knobs require a restart to take effect e.g. \textit{shared\_buffers} in PostgreSQL, so during database tuning, it is necessary to repeatedly restart the database. However, some knobs could be updated online, making it possible to tune knobs online without restarting if we only tune on these knobs~\cite{towards_safe, cdb_tune}. In this paper, we do not distinguish whether these knobs need a DBMS restart, and uniformly apply changes by restarting the DBMS. 

\textbf{ii).} We assume that the database controller has the privilege to monitor the environment information of production database, while Ottertune~\cite{van_aken_automatic_2017} also adopts the same strategy. And recent works~\cite{butrovich_tastes_2022,amazon_web_service} have allowed users to obtain runtime behavior of the system at a lower cost. In our code, the metric collector is pluggable, enabling users to insert any performance monitoring tool without affecting the overall system workflow. And for simplicity, we dismiss the cost of monitoring.

\textbf{iii).} Since modeling workload with database runtime metric is proven to be effective in Ottertune. In this paper, our WS aims to make the synthesized workload as similar as possible to the target workload at the metric level. However, in practice, due to the complexity of basic workloads, using only open-source workloads may not be sufficient to fully characterize the behavior of target workloads, so users can provide their own workloads to accomplish WS. Moreover, how to synthesize workloads more accurately is beyond this research, and we left it for future work. 

\textbf{iv).} Since every database tuning process can be recorded and becomes part of the Experience Repository, we assume that the repository contains sufficient experience for knowledge transfer.

In the following sections, we will introduce the detailed design and implementation of each part shown in Fig.~\ref{fig:overview}.
\section{WORKLOAD SYNTHESIS}\label{sec:workload}

\begin{figure}\label{fig:EET}
\centering
\includegraphics[width=3in]{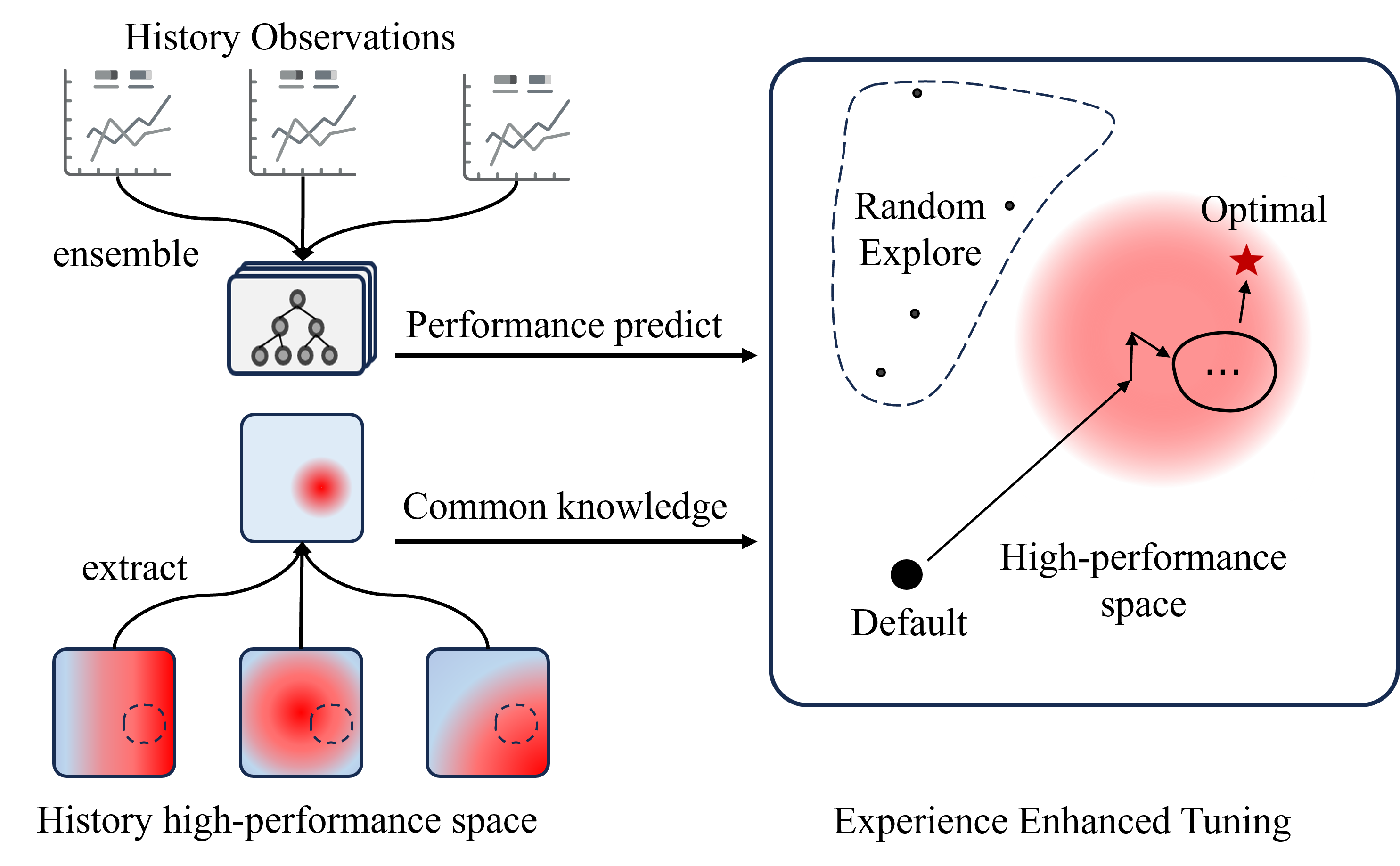}
\caption{Toy Example of EET}
\label{fig:EET}
\end{figure}

Since collecting workload information can lead to data leakage. If we directly tune the \textsc{Production DB}, the bad configurations recommended and frequent updates of configuration will likely cause a decline in database performance, during tuning. Hence, the necessity arises for a synthesized workload that can simulate current workload characteristics. Then the optimizer only needs to tune on the synthesized workload running on the \textsc{Cloned DB}, and subsequently deploy the well-performing configurations to the \textsc{Production DB}. This approach demonstrates two advantages. On the one hand, it can stably reproduce the workload characteristics 
required for the optimizer. One the other hand, it can avoid the performance impact on the database due to the optimizer's initial random exploration when directly tuning the \textsc{Production DB}, thus ensuring database performance stability.

The key idea of workload synthesis is to assign weights to the basic workloads and integrate them so that the integrated workload resembles the target workload at the metric level during workload execution. In this section, we will introduce the details of how to obtain the database's internal metrics and external metrics, and following that, we will describe how workload synthesis is implemented.

\subsection{Metric Collection}\label{sec:metric_collection}

\textbf{Internal Metrics: }
DBMS provide system views that accurately record logical information about the DBMS's behaviors during runtime, such as buffer size allocated, the number of block reads and hits, lock conflicts, and etc. 
Clearly, this information can effectively reflect the system resource consumption during the workload execution. In practice, the metric collector periodically sends requests to the DBMS to extract data from the system views, and finally we use the rate of change of each statistical metric as a representation of the current workload. 
In the experiment, we use the PostgreSQL database as the optimization target and select 65 numeric metrics to represent workload characteristics. These metrics involves global statistics, database-level statistics, table-level statistics, and index-level statistics, offering a comprehensive view of DBMS's performance under varying workloads.

\noindent\textbf{External Metrics: }
External metrics refer to the hardware resource consumption of a database on the host machine, such as CPU usage, memory consumption, and read/write rates. Similar to the aforementioned method, in implementation, we utilize the psutil module in Python to periodically monitor all processes of the database, obtaining the external metrics. The results observed are averaged to represent the final external metrics. We transform external metrics into high-dimensional vectors and concatenated with internal metrics to form the final metric results.

\subsection{Workload Generation}\label{sec:workload_generation}

Based on existing open-source workloads, we have collected the open-source workloads, such as ycsb, tpcc, twitter and etc. Each benchmark is composed of different transaction types and each transaction has different behavior. For example, \textit{ReadRecord} and \textit{InsertRecord} are two of YCSB's transaction types, and obviously they are designed to test the read and write performance of the database. And for each basic workload $\omega_b$, we can collect characteristics of every transaction of $\omega_b$. Therefore, if $\omega_b$ has $k$ transaction types, we can derive $k$ basic workloads from $\omega_b$, serving as inputs for the workload generator. 

Subsequently, we view the synthesis of workloads as a linear programming problem, assuming that we derive $n$ basic workloads from the open-source workloads. Thus, the optimization objective of this problem can be formalized as ~\autoref{alg:workloadSyn}. 
Based on this, in our experiments we specify that only a subset of $m$ basic workloads are selected to compose the synthetic workload. Without this constraint, there would be some workloads with relatively small weights, leading to troubles in workload synthesizing. Therefore, we extend ~\autoref{alg:workloadSyn} to ~\autoref{alg:workloadComp}. We use $\Gamma$ as a regularization term to constrain the total number of workloads selected by the algorithm. $\mathbb{I}$ is an indicator function that returns 1 if $(0 < \lambda_i)$ is true, or 0 otherwise.

\begin{figure}
\centering
\includegraphics[width=3in]{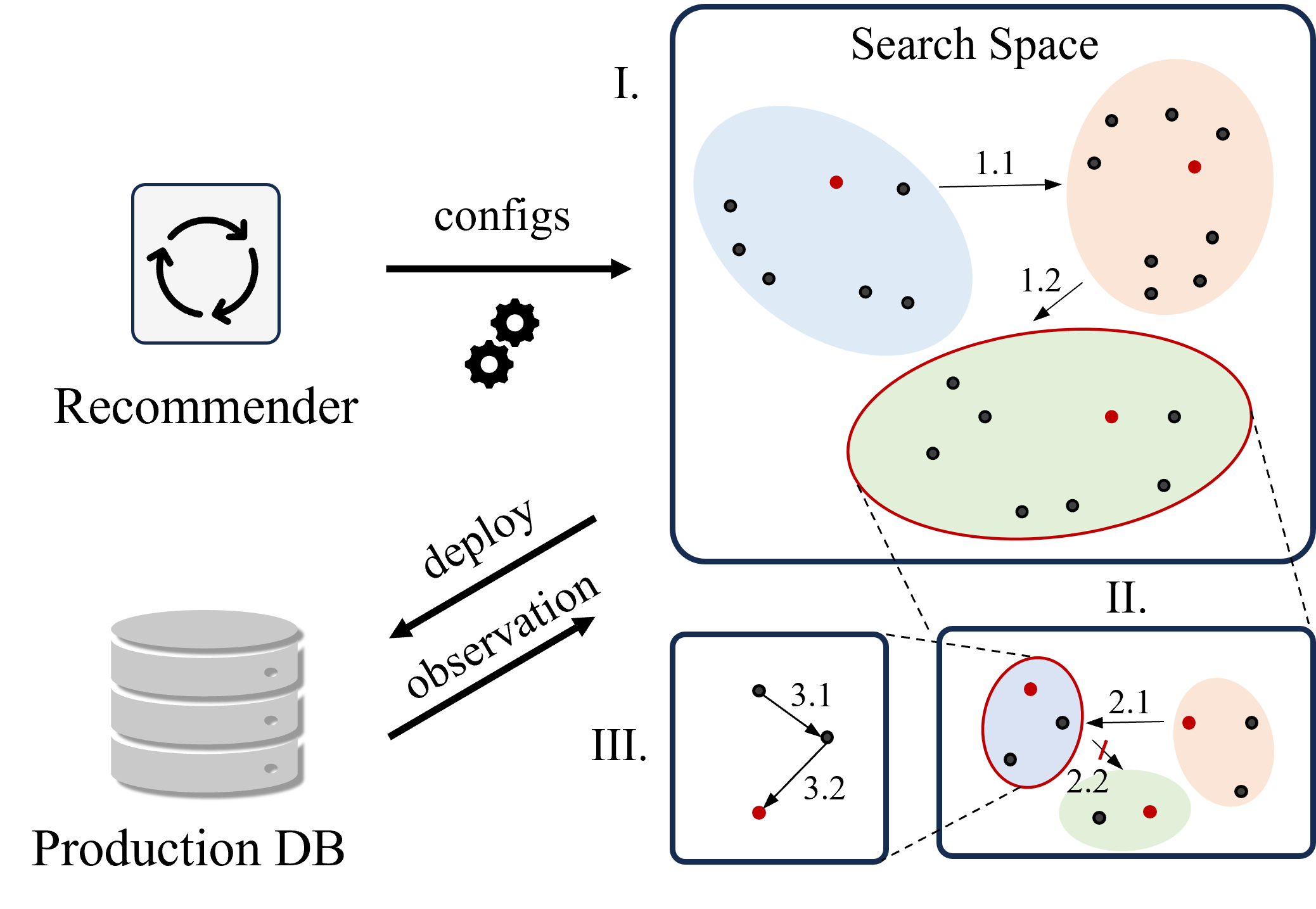}
\caption{Toy Example of Configuration Selection}
\label{fig:strategy}
\end{figure}

\begin{equation}
\label{alg:workloadComp}
\left\{ \begin{aligned}
    &\lambda^* = \mathop{\arg\max}_{\lambda \in \Lambda} \left(\text{Similarity} \left( \sum_{i=1}^{k} \lambda_i \cdot \omega_i, \omega_t \right) - \Gamma\right), \omega_i \in \Omega \\
    &\Gamma = \text{MAXINT} * \left[ \sum_{i=1}^{k}\mathbb{I} \left( 0 < \lambda_i \right) - m \right]^2\\ 
    &\text{s.t. } 0 \leq \lambda_i \leq 1 , \sum_{i=1}^{n} \lambda_i = 1, \omega \in \Omega
\end{aligned} \right.
\end{equation}

\section{EXPERIENCE ENHANCED TUNING}\label{sec:tuning}

Some current works~\cite{too_many, ibtune, hunter}, select a subset of knobs to tune based on their importance, thus pruning search space. However, these methods, while reducing the dimensions of tuning, will dismiss the interaction between knobs and potentially leading to local optima. 

In this section, we will introduce the method of tuning based on common knowledge transferring. The basic idea is that, for historical tuning tasks of various workloads, we often observe that the ranges of certain excellent configuration settings cluster within a relatively narrow band. For example, setting the \textit{shared\_buffers} parameter of a PostgreSQL database too low can prevent the database from fully utilizing caching to speed up transaction processing. If it is set too high, approaching or exceeding the physical memory size of the host machine, the database may fail to start. Therefore, it is unnecessary for the optimizer to explore excessively large or small values. In most scenarios, setting the \textit{shared\_buffers} parameter to range between 50\% to 75\% of the host machine's physical memory size can maintain good database performance, therefore this can serve as common knowledge to guide optimizer's tuning in new scenarios. For our method, the aim of common knowledge extraction is to identify those high-performance subspace. If the important knobs are set appropriately, the configuration is likely to fall within this high-performance space. 

With the constraint of common knowledge, the search space could be pruned and avoid the trap of local optima due to knob selection. Next, we will first introduce how we extract common knowledge in ~\autoref{sec:common_knowledge}, and then in ~\autoref{sec:EESMAC} we will detail how to leverage this common knowledge to guide the optimization of the SMAC algorithm.



\subsection{Common Knowledge Extraction}\label{sec:common_knowledge}

In practice, even though the workload execution of databases varies widely, we find that under optimal configurations, the range of values for certain knobs does not change much. For example, the memory size allocated to a database cannot exceed the memory of the operating system, otherwise, the database startup will fail. In some AP scenarios, allowing the database to generate query plans with nested loops might lead to a sharp increase in resource consumption for some slow queries, hence, in many cases, DBAs configure relevant knobs to prevent the database from generating query plans that include nested loop operators. Therefore, the setting of the knob's value range affects the space exploration of the database. As a result, we proposed an algorithm for common knowledge extraction, which aims to identify those common high-performance subspaces from historical experience, thus reducing some unnecessary hyperparameter exploration.

The specific algorithm is described in \autoref{alg:commonExtract}. To identity the high-performance configuration space, users can specify a high-performance threshold (line 4), and configurations exceeding this threshold are marked as high-performance configuration (lines 6-9). Then, machine learning is used to separate those high-performance spaces (line 11), and in the experiments, we employed Support Vector Machines (SVM) to extract these subspaces. These subspaces will be used to guide the optimizer exploration in \autoref{sec:EESMAC}. 

\begin{algorithm}[t]
\DontPrintSemicolon
\KwIn{Experience Repository $\mathcal{E}$, Percentile $\eta$}
\KwOut{Common Knowledge Predictors (CKPs) $\mathcal{M}$}
\caption{Common Knowledge Extract}\label{alg:commonExtract}

$ \mathcal{M} = empty\_list()$\\
\For {$e, \omega \in \mathcal{E}$}
{
    $train\_data = empty\_list()$\\
    $threshold = \eta\_percentile(e)$\\
    \For {$ config, perf \in e$}{
        \If {$perf > threshold$}{
            $train\_data.add((config, 1))$\\
        }
        \Else{
            $train\_data.add((config, 0))$
        }
    }
    \text{//fit a model on train\_data}\\
    $m = fit(train\_data)$\\
    $\mathcal{M}.add(m)$\\
}
\Return{$\mathcal{M}$}.\;
\end{algorithm}

\subsection{Experience Enhanced SMAC}\label{sec:EESMAC}
Bayesian optimization has the ability to balance exploration (the predictive mean of the surrogate model for the exploration points) with exploitation (the predictive variance of the surrogate model for the exploration points).
When faced with new scenarios, current studies\cite{shen_rover_2023, hunter}  utilize the surrogate models of history tasks and integrate them into an emsembled model, and the exploration and exploitation information are given by the emsembled model.
However, in new scenarios, we hope bayesian optimization can find near-optimal configurations as early as possible. Thus, the exploration part of historical tasks becomes less significant, as we prefer to find configurations that perform well stably in the early stage, rather than allocating budget to gamble on uncertain configurations, which may has a small chance of allowing the optimizer to find a superior configuration space, but likely to waste time and resources. Moreover, utilizing only the exploitation of historical tasks' surrogates poses challenges since different workloads have varying sensitivities to configuration changes, coupled with the uncertainty brought by performance testing. This makes it difficult for knowledge-transfer-based BO algorithms to ensure that the early exploration points significantly enhance database performance in new scenarios. 

Therefore, to increase the confidence of finding near-optimal configurations early on based on historical knowledge transfer, we embed the common knowledge described in \autoref{sec:common_knowledge} into the BO algorithm to achieve this purpose. In practice, we use the SMAC algorithm, a variant of the BO algorithm, for knowledge transfer. Choosing SMAC is primarily for the following reasons: i) According to \cite{zhang_facilitating_2022}, the SMAC algorithm has shown impressive performance in database knob tuning, indicating its strong potential in this domain. ii) The SMAC algorithm requires sampling a large number of points and then ranking them when calculating the acquisition value, which conveniently aligns with acquiring common knowledge by predicting whether the sample points fall into a high-performance space. iii) To avoid the optimizer getting stuck in local optima, we employ a decay strategy to reduce the importance of historical knowledge, and the method of calculating the acquisition value in SMAC offers convenience for this approach. Below, we will introduce the implementation of the EESMAC algorithm. The overall process of the algorithm is shown in ~\autoref{alg:EET}. Next, we will introduce the specific implementation of Experience Enhanced SMAC (EEMSAC). 

Initially, we employ a metric collector to gather the characteristics of the current execution workload. Following that, we match the current workload with similar historical workloads based on their characteristics (lines 2-3). Subsequently, weights are assigned to each matched workload based on their similarity to the current workload. Intuitively, the more similar a historical workload is to the current one, the more valuable it is for reference in the current tuning (line 4). These weights are then utilized to calculate the acquisition value, which is detailed in ~\autoref{alg:EEA}. In the initial phase, facing a new scenario without any prior knowledge, we use a ratio $\kappa$ as the exploration factor. During this exploratory phase, the optimizer will solely rely on historical experiential knowledge (lines 11-15). After the initial phase, we adopt a decay strategy, gradually reducing the weight of historical knowledge to balance the knowledge of new scenarios with experiential knowledge (line 20). As shown in ~\autoref{eq:hist_curr_trade}.

\begin{equation}
\begin{aligned} \label{eq:hist_curr_trade}
& ac\_value = \zeta * knowledge_{hist} + (1-\zeta) * knowledge_{curr}\\
\end{aligned}
\end{equation}

$\zeta$ represents the weight of historical knowledge, which is initially set to 1 and decreases with the number of exploration iterations, allowing the optimizer to focus more on the exploitation of currently observed data (line 21). The specifics of the acquisition function algorithm are detailed in ~\autoref{alg:EEA}, and as illustrated in lines 9-10, both transferred knowledge and current observations are taken into account for optimization. 

Due to the exploitation of exploration for historical knowledge when computing the acquisition function value, the time complexity has been increased compared to the original version. In SMAC, a Random Forest~\cite{breiman2001random} is employed as the surrogate model. Suppose that the number of knobs to be tuned is $q$, and the number of configurations have already explored is $n$. For each training session, the training time complexity of the Random Forest is $O(n \cdot q \cdot \log n)$, and for prediction, the time complexity for each point is $O(\log n)$. In SMAC, during each iteration, multiple samples need to be drawn to select the point with the highest acquisition value for observation. Assuming the number of samples is $m$, the predictive time complexity for each iteration round is $O(m \cdot \log n)$. At the $(n+1)$-th iteration, the total time complexity of SMAC would be $O( n \cdot q \cdot \log n) + O(m  \cdot \log n) = O(q \cdot n \cdot \log n + m \cdot \log n)$. For the computational complexity of history models, supposing we utilize $h$ history tasks, all the models in $\mathcal{M'}$ and $\mathcal{M}_p$ need to perform a prediction for each sample. Suppose that we use SVM(Support Vector Machine) with linear kernel for $\mathcal{M'}$, and each model in $\mathcal{M}_p$ is a random forest, thus the computational complexity for history knowledge transfer is $O(m \cdot q \cdot n)$ and $O(m  \cdot \log n)$. The overall complexity comes to $O( (q \cdot n + m) \cdot \log n) + h \cdot( O(m \cdot q \cdot n) + O(m \cdot \log n)) = O(q n \log n + hmqn + hm\log n)$. In practice, the overall time complexity is primarily determined by the number of samples ($m$) and the number of iterations ($n$).

\subsection{Example}\label{sec:eet_example}

~\autoref{fig:EET} intuitively demonstrates the tuning process of EET. Initially, the optimizer can extract the high-performance space to constrain the direction of knob tuning exploration at a coarse degree, by exploiting history experiences. To exploit historical experiences at a finer degree, we adopt an ensemble strategy, integrating well-fitted history models to predict the performance of sampled points. In ~\autoref{fig:EET}, assuming \textit{Default} as the starting point, the exploration would be more inclined towards the high-performance spaces extracted in history. Meanwhile, the optimizer employs an $\epsilon$-greedy strategy for random exploration to ensure avoiding local optima as much as possible in the early stage. Due to the decay strategy, the optimizer will eventually convergence to global optima.

\begin{algorithm}[t]
\DontPrintSemicolon
\KwIn{Experience Repository $\mathcal{E}$, Decay $\gamma$, Init Ratio $\kappa$\\
\hspace{0.53cm} Transfer Quantity $N$, Max Iteration $\mathcal{I}$, Random Ratio $\epsilon$ \\}
\KwOut{Best Configuration $x^*$}
\caption{Experience Enhanced SMAC}\label{alg:EET}
$\mathcal{P} = \{\}$\\
$\omega_t = \text{Collect metrics from workload.}$\\
$\mathcal{E}' = \text{Top N most similar experiences based on $\omega_t$ from }\mathcal{E}$\\
$\mathcal{S} = \text{normalize}\left([\text{similarity}(\omega_i, \omega_t) | \omega_i \in \mathcal{E'}]\right)$\\
\text{//obtained by Algorithm 3.}\\
$\mathcal{M}' = \text{common\_knowledge\_extract}(\mathcal{E}')$\\
\text{//history performance predict models.}\\
$\mathcal{M}_p = [fit(e_i)|e_i \in \mathcal{E'}]$\\
$m_s = surrogate\_model()$\\
\While {$|\mathcal{P}| < \mathcal{I}$}
{
    \If{$|\mathcal{P}| < \kappa * \mathcal{I}$}{
        \text{//calculated by Algorithm 5.}\\
        $p = p_i \text{ maximizes } f(p_i, \mathcal{M'}, m_s, \mathcal{M}_p, \mathcal{S}, 1)$\\
        $\mathcal{P}.add((p, Observation(p)))$\\
        \text{continue}\\
    }
    \If{$random() < \epsilon$}{
        $p = random\_sample()$\\
    }
    \Else{
        $\mathcal{M}_s = fit(\mathcal{P})$\\
        $\zeta = max\left(1-\gamma * (|\mathcal{P}| - \kappa * |\mathcal{I}|), 0\right)$\\
        $p = p_i \text{ maximizes } f(p_i, \mathcal{M'}, m_s, \mathcal{M}_p, \mathcal{S}, \zeta)$\\
        $\mathcal{P}.add((p, Observation(p)))$\\
    }
}
$p^* = \text{Best observed configuration in }\mathcal{P}$\\
\Return{$\left(p^*, \mathcal{P}\right)$}.\;
\end{algorithm}

\begin{algorithm}[t]
\DontPrintSemicolon
\KwIn{Unseen Point $p$, CKPs $\mathcal{M}'$, Surrogate Model $m_s$,\\
\hspace{0.33cm} Performance Predictor $\mathcal{M}_p$, Similarity $\mathcal{S}$, Trade Off $\zeta$}
\KwOut{Score $s$}
\caption{Experience Enhanced Acquisition Function}\label{alg:EEA}
$votes = [m_i.predict(p) | m_i \in \mathcal{M}']$\\
$perfs = [m_j.predict(p) | m_j \in \mathcal{M}_p]$\\
\text{//comprehensive knowledge transfer}\\
$s = votes \cdot \mathcal{S}^T + perfs \cdot \mathcal{S}^T$\\
\If{$\zeta = 1$}{
    \text{// $m_s$ is not trained, so return directly.}\\
    \Return{$s$}\\
}
\Else{
    $\mu, \sigma = m_s.predict(p)$\\
    $s = \zeta * s + (1-\zeta) * (\mu + \sigma)$\\
}
\Return{$s$}.\;
\end{algorithm}

\section{CONFIGURE REPLACEMENT STRATEGY}\label{sec:strategy}

In practical scenarios, there is a gap between synthetic and actual workloads due to the ability of workload synthesis, which means configuration recommended by the optimizer might still cause database performance regression, leading to decreased workload execution efficiency and resulting in unnecessary losses. Therefore, the safety of knob configuration becomes particularly important. Given that current methods struggle to guarantee the new database configurations will outperform the original settings~\cite{towards_safe}, we proposed a unique database configuration replacement strategy. It employs a mechanism based on confidence ranking and rollback, aiming to ensure that the new database configuration recommended is superior to the original one as much as possible. If any performance regression occurs, the database configuration will be rolled back to minimize the impact on workload.

\subsection{Configuration Selection}

Based on the output of ~\autoref{alg:EET}, we obtain all the observed points $\mathcal{P}$ for the synthetic workload on the \textsc{Clone DB}. Hence, we select the top-k configurations with best performance as the configuration candidates set $\mathcal{P}' = \{p_1, p_2, ..., p_k\} \subset \mathcal{P}$. 
An intuitive approach is to greedily try each candidate configuration one by one, selecting the one with the best actual performance, or stopping when a desired performance improvement is achieved. However, this method can lead to unnecessary attempts. For example, two configurations that are very close in the configuration space are likely to exhibit very similar performance. If one configuration performs poorly on the \textsc{Production DB}, the other is also likely to lead to poor database performance, making the trial of another configuration unnecessary. 

Therefore, we propose a recursive-cluster configuration replacement strategy, with the specific implementation process as ~\autoref{alg:confSelection}: 
\textbf{i). Line 1-4:} First, according to $\mathcal{P}'$, we cluster the configuration items to obtain $k$ clusters, where the configurations in close proximity within the configuration space will be assigned to the same cluster.
\textbf{ii). Line 5-14:} Then, we find the globally optimal point and deploy it to the \textsc{Production DB}. If performance regression occurs, discard all configurations within the cluster that contains this point, until a configuration that enhances performance is identified as $p$.
\textbf{iii). Line 15-21:}  After finding a point that improves performance, begin exploring the set of configurations most similar to it and iteratively update the point $p$ until a performance decline is observed. This likely indicates that the exploration has passed the region of the global optimum, thus an early stop is warranted. This can also reduce the cost associated with changing database knobs.
\textbf{iv). Line 22:}  We recursively explore within the optimal set until the stopping criteria are satisfied. 

Based on ~\autoref{alg:confSelection}, we assume that $n$ configurations are selected from those already explored by the optimizer for configuration selection. In the worst-case scenario, the first iteration requires observe $\mathcal{K}$ configurations, and each subsequent iteration only needs to observe $\mathcal{K}-1$ configurations. Each iteration re-clusters a set of configurations, so on average, $O(\log n)$ iterations are needed; therefore, a total of at most $O(\mathcal{K}\log n)$ database configuration updates are required to determine the optimal configuration.

\subsection{Example}
~\autoref{fig:strategy} shows a toy example of configuration selection. Initially, the Recommender provides a set of configurations that perform well under synthetic workloads, which are clustered according to their positions in the hyperparameter space. Subsequently, a recursive trial is conducted. In steps 1.1 and 1.2, it is discovered that new clusters outperform the old ones, hence the last visited cluster is selected for the second round of clustering and exploration. In step 2.1, the new cluster outperforms the old one, while in step 2.2, a performance regression occurs, leading to the termination of exploration. The cluster with the highest performance is selected for the third exploration. Through the exploration in steps 3.1 and 3.2, the optimizer identifies the currently optimal configuration as the final result.

\begin{algorithm}
\DontPrintSemicolon
\KwIn{Configuration Candidates $\mathcal{P}'$, Max Depth $\mathcal{I}$,\\ Current Performance $perf_c$, Threshold $\epsilon$, Cluster Number $\mathcal{K}$\\
}
\KwOut{Final Configuration $p^*$}
\caption{Recursive Configuration Selection}\label{alg:confSelection}
$i = 0$\\
\While{$i < \mathcal{I}$ and $|\mathcal{P}'| > 1$}{
    $i = i + 1$\\
    $clusters = KMeans(\mathcal{P}', \mathcal{K})$\\
    \While{$\mathcal{P'} \neq \emptyset$}{
        $p = \text{Best point in } \mathcal{P}'$\\
        $perf' = Observation(p.x)$\\
        \If{$\frac{perf' - perf}{perf} < \epsilon$}{
            $\mathcal{P}' = \mathcal{P}' - (cluster | p \in cluster)$\\
            $continue$\\
        }
        $break$\\
    }
    \If{$\mathcal{P} = \emptyset$}{
        \text{// Restart knob tuning.}\\
        \Return{None}
    }
    \While{Not all clusters visited}{
        \text{// $Cluster(p)$ returns the cluster where p in.}\\
        $p' =$ Best point in cluster closest to $Cluster(p)$\\
        $perf'' = Observation(p.x)$\\
        \If{$perf'' < perf'$}{
            $break$\\
        }
        $perf',p = perf'',p'$\\
    } 
    $\mathcal{P'} = Cluster(p)$\\
}

\Return{$p.x$}.\;
\end{algorithm}

\begin{figure}
\centering
\includegraphics[width=3in]{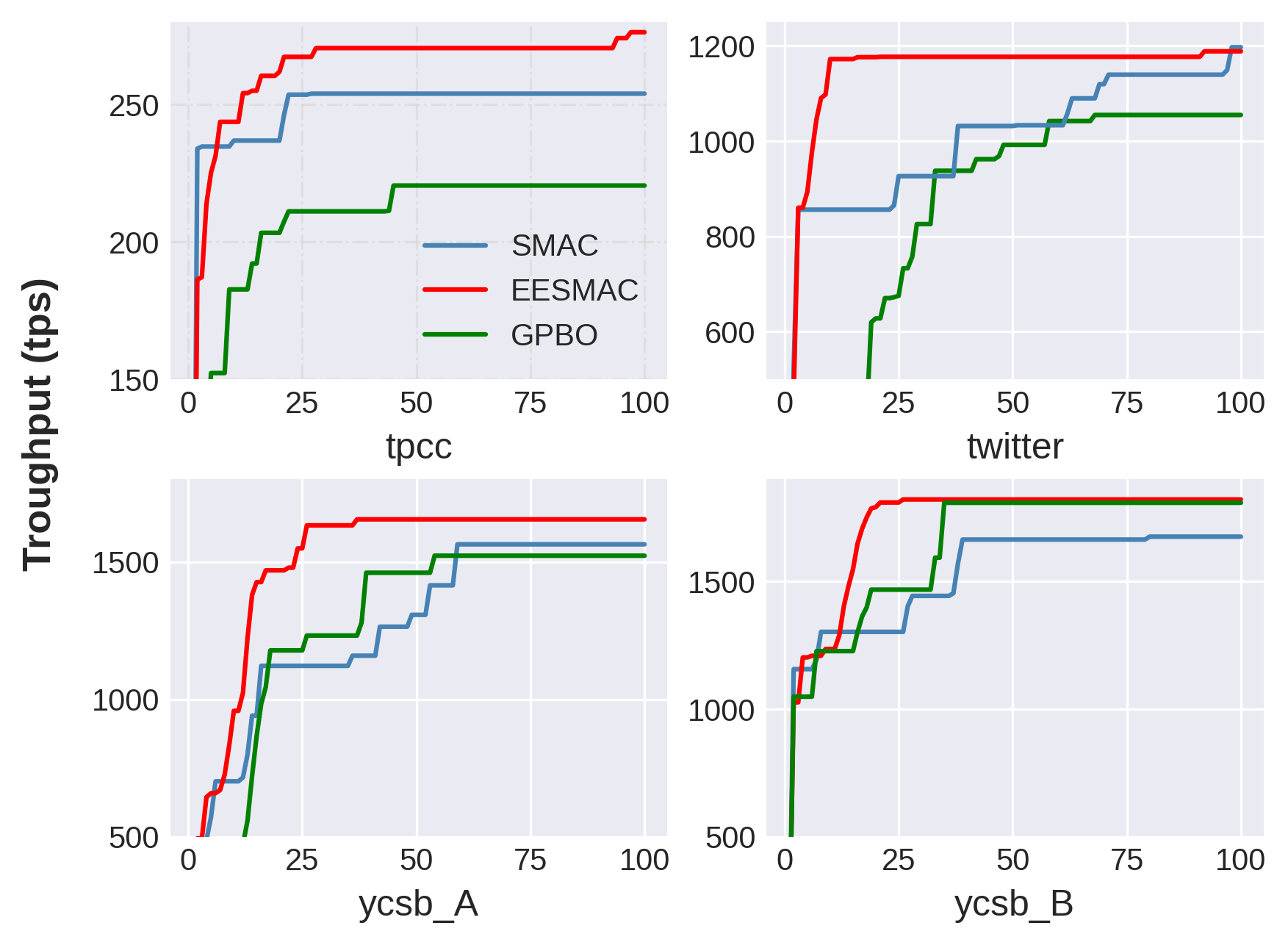}
\caption{Evaluation of Experience Enhanced SMAC}
\label{fig:eval_eet}
\end{figure}

\section{EXPERIMENTS}\label{sec:experiment}

\begin{figure*}
\centering
\includegraphics[width=6.5in]{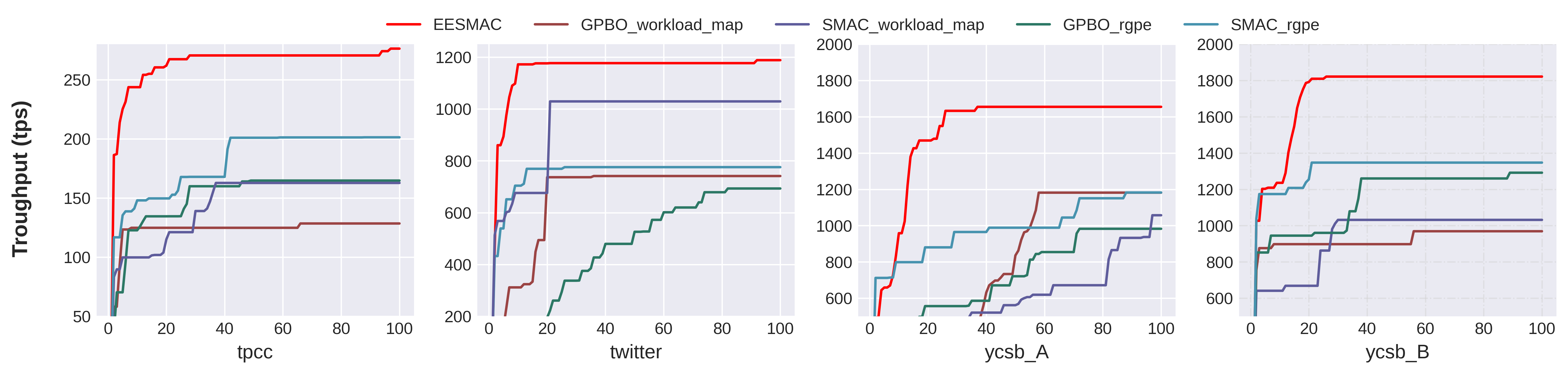}
\caption{Evaluation of Knowledge Transfer Methods}
\label{fig:tansfer}
\end{figure*}

\begin{figure}
\centering
\includegraphics[width=3in]{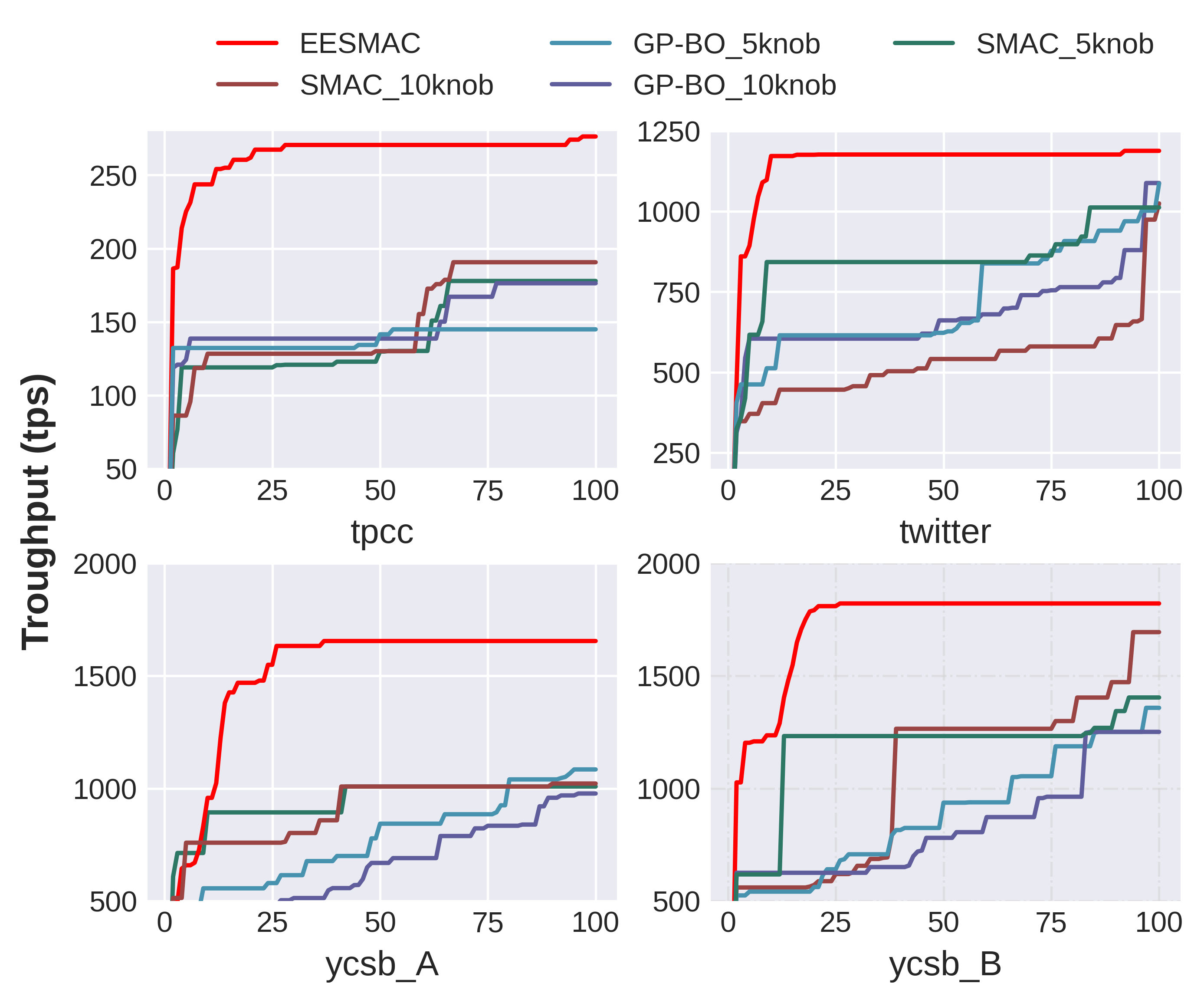}
\caption{Hyperparameter Space Reduction}
\label{fig:knobs_num}
\end{figure}

In this module, we will evaluate \sys under various workloads. We will demonstrate that our proposed Experience Enhanced SMAC accelerates the process of Bayesian Optimization (BO) in finding near-optimal configurations. Additionally, our approach significantly improves database performance with only a few configuration updates to the \textsc{Production DB}.

\subsection{Experiment Setup}
\noindent\textbf{Environment Setting.} 
To limit the hardware resources of DBMS, we deploy the DBMSs in Docker\cite{docker} to constrain the resource usage of the DBMS. We restrict Docker's CPU count to 4 and RAM size to 8GB, with external storage of a RAID5 disk array for data redundancy. The DB Controller are isolated from the DBMS environment. It interacts with the DBMS through socket to send workloads to them. And during database configuration updates, the DB Controller uses SSH to upload and replace the existing database configuration file, followed by restarting the database to apply the new settings for the knobs. 

For knob tuning, we conduct all experiments on PostgreSQL v13.13. We have selected 30 tunable knobs from the database configurations that include settings related to memory, optimizer, background processes, and query executor and etc. These knobs allow for a comprehensive optimization of database performance. 

We aim to optimize the throughput of the workload and for each workload, the database is first warm up for one minute, then execute the workload for five minutes to ensure the stability of the performance results in every iteration. After each optimization step, the optimizer will report the best configuration found and its performance so far for the current iteration. 
Each tuning session consists of 100 iterations. Since Bayesian optimization cannot perform cold starts, for the initial 10 iterations, we employ a random sampling method to generate 10 configurations and then explore them to obtain observation, which enables the surrogate model to construct an initial probability distribution.

\noindent\textbf{Workload Setting.}
We would like to extend our gratitude to benchbase\cite{benchbase} for designing the all-in-one database benchmarking tool. In experiment implementation, we employed the widely-used benchmarks TPCC, YCSB\_A, YCSB\_B, and the Twitter dataset to conduct a comprehensive performance test of our approach. And for the properties of benchmarks, the number of tables, the total number of columns, and the proportion of read-only transactions to the total transactions for each benchmark are presented in ~\autoref{tab:workload_prop}. YCSB-A and YCSB-B represent two variants of the YCSB workload with different ratios of query configurations, where YCSB-A is a balanced read/write workload, while YCSB-B is read-heavy. On the other hand, TPC-C is a traditional OLTP workload that uses 9 tables and 92 different columns to simulate the querying scenarios of an online shopping cart.And Twitter is a web-oriented workload that utilizes public traces to simulate the characteristics of a micro-blogging platform. In the experiments, we set the size of all datasets to 10GB, exceeding the 8GB RAM of their respective containers. Furthermore, we configured an unlimited send rate to capture the maximum throughput of the workloads. For the experience repository, we generate with random ratio of transactions for each workloads, and they are used for knowledge transfering through the entire experiments.

\noindent\textbf{Optimizers.}
In our experiments, we employed two variants of Bayesian optimization: GP-BO, which uses Gaussian Processes as the surrogate model for Bayesian optimization, and SMAC, which utilizes Random Forests as its surrogate model. Both Bayesian optimization methods have demonstrated superior performance in many cases\cite{zhang_facilitating_2022}. Furthermore, we leveraged the code framework of DBTune \cite{DBTune} to implement our algorithms.

\begin{table}[!t]
    \caption{Workload Properties.}
    \resizebox{0.35\textwidth}{!}{
    \renewcommand{\arraystretch}{1}
        \begin{tabular}{clccc}
        \toprule
         Workload       &       &   \multicolumn{1}{c}{Tables} & \multicolumn{1}{c}{Columns}  & \multicolumn{1}{c}{Read-Only}    \\  
        \midrule
        \textbf{YCSB-A} &       & 1 &   11      &   50\% \\
        \textbf{YCSB-B} &       & 1 &   11      &   95\%   \\
        \textbf{Twitter}&       & 5 &   18      &   1\%    \\
        \textbf{TPCC}   &       & 9 &   92      &   8\%     \\
        \bottomrule
        \end{tabular}
    }
    \label{tab:workload_prop}
\end{table}

\begin{table*}[!t]\footnotesize
    \caption{iterations to achieve (ratio $\times$) the best performance found by EESMAC }
    \renewcommand{\arraystretch}{1}
    \begin{tabularx}{0.85\linewidth}{cccccccc|ccccc}
        \toprule
            & & \multicolumn{6}{c}{SMAC} & \multicolumn{5}{c}{GP-BO} \\
            \hline
            & & \multicolumn{11}{c}{speed up method}\\
        \midrule
            Dataset & ratio & Ours &W.M. & RGPE & 5 knobs & 10 knobs & None  & W.M. & RGPE & 5 knobs & 10 knobs & None \\
            \hline
            YCSB-A  & 0.7  & \textbf{13}   & -    & 88 (6.8)   & -       & -        & 36 (2.8)   & 58 (4.5)  & -    & -       & -        & 18 (1.4)  \\
                    & 0.9  & \textbf{24}   & -    & -    & -       & -        & 59 ( 2.6)   & -    & -    & -       & -        & 54 (2.3)   \\
            YCSB-B  & 0.7  & 12   & -    & 21 (1.8)  & 90  (7.5)    & 76   (6.3)    & \textbf{8} (0.67)    & -    & 89 (7.4)  & 97 (8.1)     & -        & 16(1.3)  \\
                    & 0.9  & \textbf{16}   & -    & -    & -       & 94   ( 5.9)    & 39 ( 2.4)   & -    & -    & -       & -        & 35 (2.2)  \\
            TPCC    & 0.7  & 4    & -    & 42 (11.5)  & -       & -        & \textbf{2} (0.5)    & -    & -    & -       & -        & 16 (4)  \\
                    & 0.9  & \textbf{12}   & -    & -    & -       & -        & 22  (1.8)  & -    & -    & -       & -        & -    \\
            Twitter & 0.7  & \textbf{3}    & 21 (7)  & -    & 84 (28)     & 97  (32)     & \textbf{3} (1)    & -    & -    & 92  (30.7)    & 92  (30.7)     & 33 (11)  \\
                    & 0.9  & \textbf{8}    & -    & -    & -       & -        & 63 (7.9)   & -    & -    & 100  (12.5)   & 97 (12)      & -    \\

        \bottomrule
    \end{tabularx}
    \label{tab:exp}
\end{table*}

\subsection{Evaluation of EESMAC}

\noindent\textbf{Compare with non-tansfer methods.}
We compared EESMAC with GP-BO and SMAC, and the experimental results, as shown in ~\autoref{fig:eval_eet}, indicating that our method can converge more quickly and identify more optimal database configurations in the early stages. In contrast, GP-BO and SMAC require a greater number of iterations to achieve the same performance level as EESMAC. Especially, GP-BO exhibits poorer performance on the TPCC and Twitter datasets.

\noindent\textbf{Compare with tansfer methods.}
We selected two methods of knowledge transfer for Bayesian optimization for comparison: workload mapping, and RGPE \cite{feurer2018scalable}. Workload mapping, proposed by OtterTune, matches the current workload with the most similar historical workload based on the Euclidean distance of database metrics. It then reuses these historical observations to construct a surrogate model for exploration. For RGPE, or ranking-weighted Gaussian process ensemble, it weights the base GP models of similar historical tasks with distinguishable weights. These weights are assigned based on the similarity between the current and historical tasks. 

In this experiment, we combined GP-BO and SMAC algorithms with workload mapping and rgpe to compare their transferability performance with EESMAC. We observed that the knowledge transfer methods based on workload mapping and RGPE did not produce good results. In all cases, they could not find better configurations than EESMAC. This is mainly due to the fact that the experiences in the experience pool did not generalize well to new scenarios. This limitation arises because the historical surrogate models are mainly designed to fit database configurations and their corresponding performance metrics. With only a limited amount of data available (100 iterations), surrogate models struggle to accurately fit high-dimensional data. Furthermore, the performance curves corresponding to historical experiences may mislead optimizers into exploring suboptimal hyperparameter spaces, thereby restricting the optimizers' exploration towards better configurations. In contrast, RGPE takes a larger amount of historical experience into account, and outperforms the workload mapping method in most scenarios. while EESMAC effectively utilizes the relevant information in the high-performance spaces from historical data, enabling the optimizer to explore more freely within high-performance spaces. Additionally, the deployment of a decay strategy allows the optimizer to balance current observations, thus enabling the discovery of superior solutions.

\noindent\textbf{Hyperparameter space reduction}
The current method for pruning the hyperparameter space size involves ranking the importance of knobs and selecting the important ones for tuning. Therefore, we chose the SHAP algorithm \cite{lundberg2017unified} to filter the important knobs. The SHAP algorithm is used to interpret the relationship between performance and features, and the work in \cite{zhang_facilitating_2022} has shown that using this algorithm for knob selection can achieve the best performance. Our method utilizes the extraction of high-performance parameter spaces to reduce the optimizer's exploration in meaningless hyperparameter spaces. Through this experiment, we aim to demonstrate the effectiveness of our method in pruning the hyperparameter space. 
Though knob importance based method reduces the search space for optimizer hyperparameters, its ignorance of the complex interactions between knobs resulting in suboptimal spaces. Conversely, EESMAC accelerates the optimizer's convergence by extracting a high-performance space to cut the search space. Results in ~\autoref{fig:knobs_num} shows that our method has a superior performance.


In ~\autoref{tab:exp}, we use the best performance identified via the EESMAC method as a reference point, and subsequently collect the number of iterations required for different methods to achieve 70\% and 90\% of this optimal performance. In ~\autoref{tab:exp}, bolded font indicates the best-performing metric, and W.M. is short for workload mapping. For reaching 90\% of the optimal performance, our method requires the fewest number of iterations. Furthermore, in comparison to other acceleration methods, our method can achieve the same level of performance improvement with $1.8\times$ to $12.5 \times$ fewer iterations. While aiming to reach 70\% of the optimal performance, our method outperforms the most of other methods. Although it requires a greater number of iterations in the YCSB-B and TPCC datasets, it still quickly identifies configurations that meet this performance in the early stages.

\subsection{End-to-end Evaluation}

\begin{figure}
\centering
\includegraphics[width=3in]{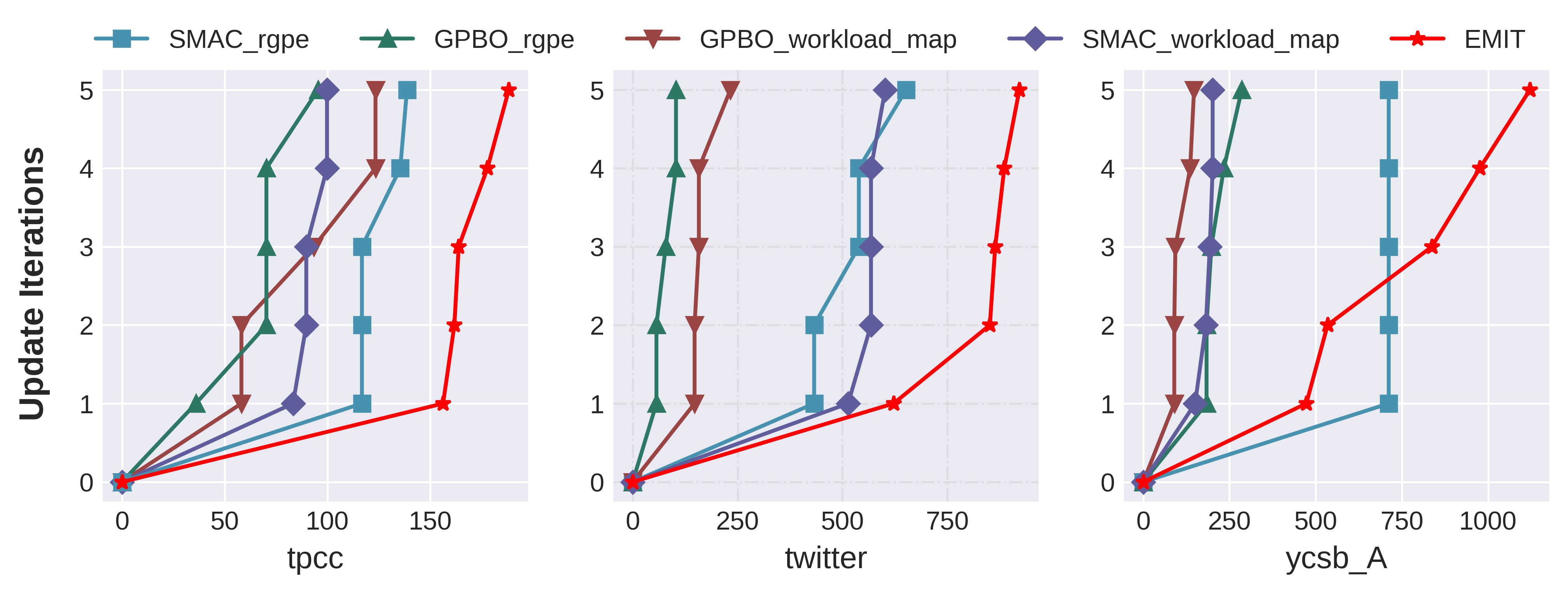}
\caption{Number of Configuration Updates}
\label{fig:updates}
\end{figure}

In this section, we conduct an end-to-end test of the system's overall pipeline. To simulate the actual scenario of workload synthesis, we select one workload for optimization while using the other two types of workloads to synthesize the target workload. For example, if we aim to optimize the Twitter workload, we first collect characteristics of the Twitter workload. Then, we use the methods described in ~\autoref{sec:workload_generation} to generate a synthetic workload with YCSB and TPC-C. After that, the optimizer tunes on the synthetic workload to derive a set of candidate configurations. After obtaining these configurations, the method mentioned in ~\autoref{sec:strategy} is employed to replace the knob settings of the \textsc{Production DB} to identify suitable knobs. The optimizer selects the top 27 performing configurations from the candidates found based on the synthetic workload. Moreover, each time we employ Kmeans for recursive search, we choose a cluster number of 3, measuring the number of iterations needed for database knob updates to achieve performance level for experimentation. The results of the experiment are shown in ~\autoref{fig:updates}. 
It can be seen from the results that our method can improve database performance by using fewer database configuration replacement steps in most cases.

\section{CONCLUSION}\label{sec:conclusion}
In this paper, we propose an efficient and micro-invasive database tuning method called \sys, which could efficiently recommend configurations for high performance, while trying to have as little intrusion and performance impact on the database as possible. To achieve the goal, we devide our system into three parts: (1) Workload Synthesis, (2) Experience Enhanced Tuning, (3) Configuration Replacement Strategy. With these parts, we could tune the database while not access the workload being executed. We test \sys on YCSB, TPCC and Twitter, and compare it with state-of-the-art tuning systems. The results show that the method can achieve performance improvements comparable to other methods with relatively fewer iterations. 

\noindent\textbf{Future Work}. In future work, we plan to investigate methodology for synthesizing workloads of arbitrary complexity and apply it to the optimization of database systems. Furthermore, we intend to explore how algorithms can autonomously extract interpretable expert knowledge from experiences, thereby enhancing the stablity and efficiency of database system tuning.


\bibliographystyle{ACM-Reference-Format}
\balance
\bibliography{main}

\newpage
\appendix
\setcounter{table}{0}
\setcounter{figure}{0}
\setcounter{section}{0}

\renewcommand{\thetable}{A\arabic{table}}
\renewcommand\thefigure{A\arabic{figure}}
\renewcommand\thesection{A\arabic{section}}

\end{document}